\def\Msun{\hbox{$\rm\thinspace M_{\odot}$}}
\documentstyle[psbox]{WORKSHOP}
\begin{document}

\title{Relativistic Disc lines\\
}

\author{A.C.~Fabian$^1$, E.~Kara$^1$, M.L.~Parker$^1$\\
%
$^1$ Institute of Astronomy, University of Cambridge, Madingley Road,
Cambridge CB3 0HA, UK \\
{\it Email: acf@ast.cam.ac.uk} 
}

\abst{Broad emission lines, particularly broad iron-K lines, are now
  commonly seen in the X-ray spectra of luminous AGN and Galactic
  black hole binaries. Sensitive NuSTAR spectra over the energy range
  of 3--78 keV and high frequency reverberation spectra now confirm
  that these are relativistic disc lines produced by coronal
  irradiation of the innermost accretion flow around rapidly spinning
  black holes. General relativistic effects are essential in
  explaining the observations. Recent results are briefly reviewed
  here.  }

\kword{black hole physics: accretion discs, X-rays: galaxies}

\maketitle
\thispagestyle{empty}

\section{Introduction}
The innermost regions around a luminous accreting black hole consist
of an accretion disc of ionized gas extending down to the Innermost
Stable Circular Orbit (ISCO) allowed by General Relativity. Material
within that radius plunges into the black hole. Differential
rotation within the disc amplifies magnetic fields creating and powering a 
compact corona above and below the centre of the disc. Hot electrons 
in the corona inverse Compton scatter soft disc photons into a hard 
X-ray power-law continuum which carries most of the radiated primary
luminosity in the X-ray band of Active Galactic Nuclei (AGN) and Hard
State Galactic Black Hole Binaries (BHB).

%
%
\begin{figure*}
\centering
\psbox[xsize=10cm]
{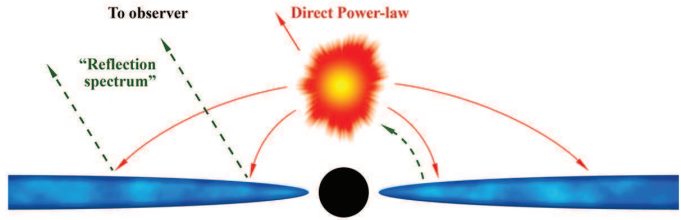}
\caption{Cartoon respresentation of the accretion disc and  black hole
with the corona above. The strong gravity causes  causes primary emission from the
corona to be bent down onto the disc. Backscattered and fluorescent
and other secondary 
emission caused by irradiation of the disc forms the reflection spectrum. }
\end{figure*}

\begin{figure*}
\centering
\psbox[xsize=10cm]
{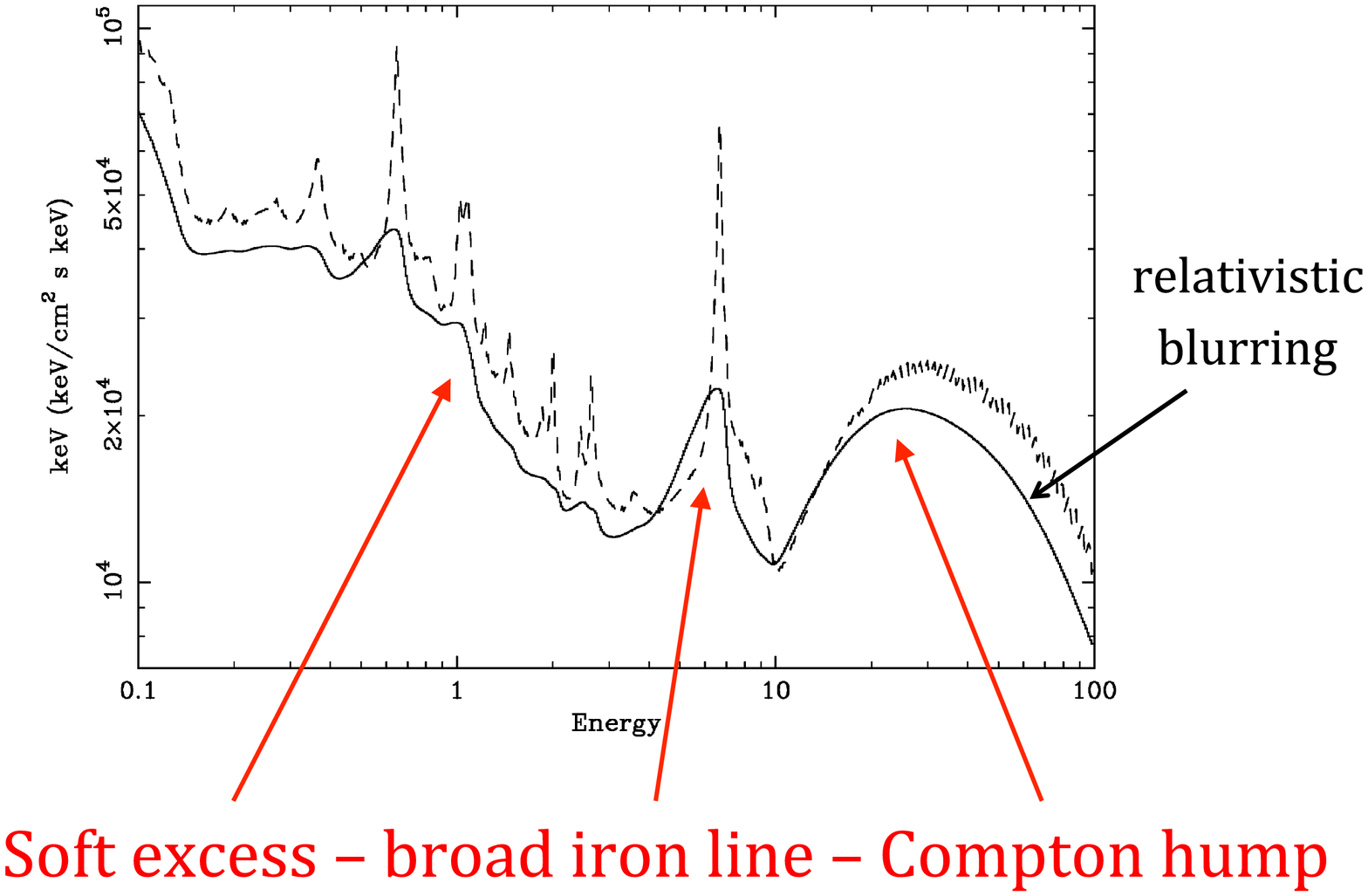}
\caption{The intrinsic reflection spectrum (dashed) is observed to be
  relativistically blurred (continuous line) by GR effects (from Ross
  \& Fabian 2005).}
\end{figure*}

Secondary emission is produced by irradiation of the accretion disc by
the primary continuum. The resulting back-scattered, fluorescent
and secondary thermal
emission (Fig.~1) is known as the {\it reflection spectrum} (even though it
does not obey the standard laws of reflection). Emission lines 
observed from the innermost parts of the disc where matter is
travelling up to half the speed of light are relativistically
broadened and skewed to the red (i.e. lower energies) by Doppler and
gravitational shifts and are shaped by relativistic beaming -- they are
relativistic disc lines. Such lines were predicted 25 years ago (Fabian et
al 1989) and first clearly seen with ASCA 5 years later (Tanaka et al
1995). 

The reflection continuum is characterized by a broad hump above 10 keV,
shaped at high energies by the Compton effect (photons lose energy
when backscattered) and at low energies by photoelectric absorption.
The line component is due to fluorescence, excitation and ionization
of the reflecting gas. Computation of the total reflection spectrum
is a non-linear process where the ionization state depends on the flux
of irradiating photons (see Ross \& Fabian 1993, 1995; Garcia et al
2014). The relativistic reflection spectrum is obtained by
relativistically blurring the whole spectrum (Fig.~2). Its
shape consists of a soft excess, a broad iron line and a Compton hump,
all of which features are now regularly observed (Fig.~3).    

Confirmation of this picture has emerged from discovery of the
expected X-ray reverberation signatures, first clearly detected from iron-L emission
(Fabian et al 2009) and then from iron-K (Zoghbi et al 2012). Following
intrinsic luminosity variations of the corona, energy
bands dominated by reflected emission lag behind bands dominated by
the primary continuum. This is a simple consequence of the longer path 
taken by the reflected signal in reaching us. 

\begin{figure*}
\centering
\psbox[xsize=9cm]
{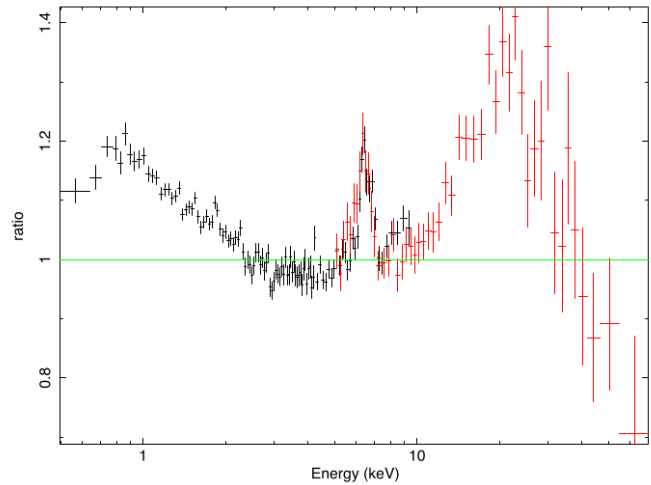}
\caption{XMM-NuSTAR spectrum of SWIFT\,J2127 (Marinucci et al 2013)
  shown as a ratio to a power-law. Note that all aspects of the
  reflection spectrum; the soft excess, the broad iron line and the
  Compton hump are seen here.}
\end{figure*}
        
\begin{figure*}
\centering
\psbox[xsize=7cm]
{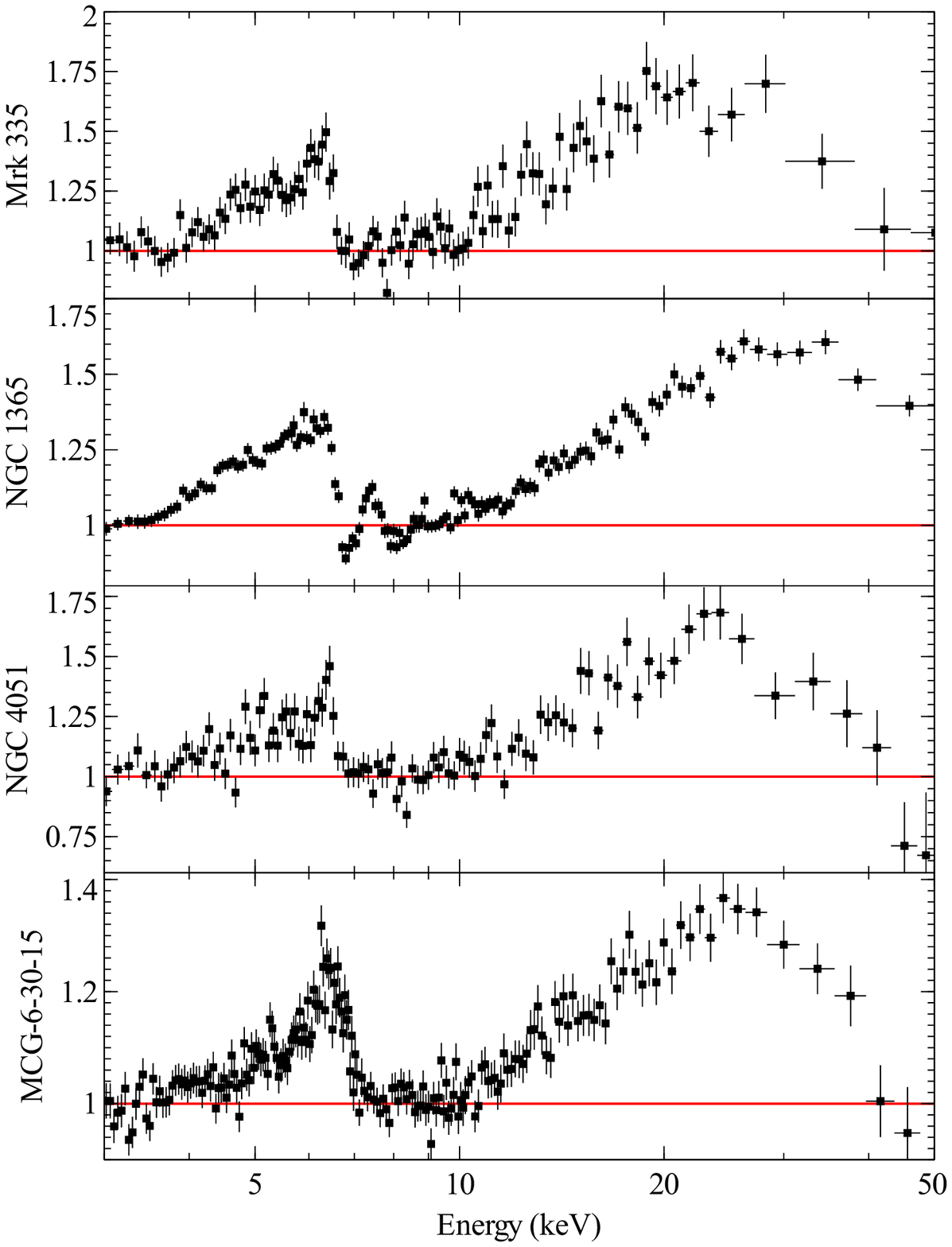}
\psbox[xsize=7cm]{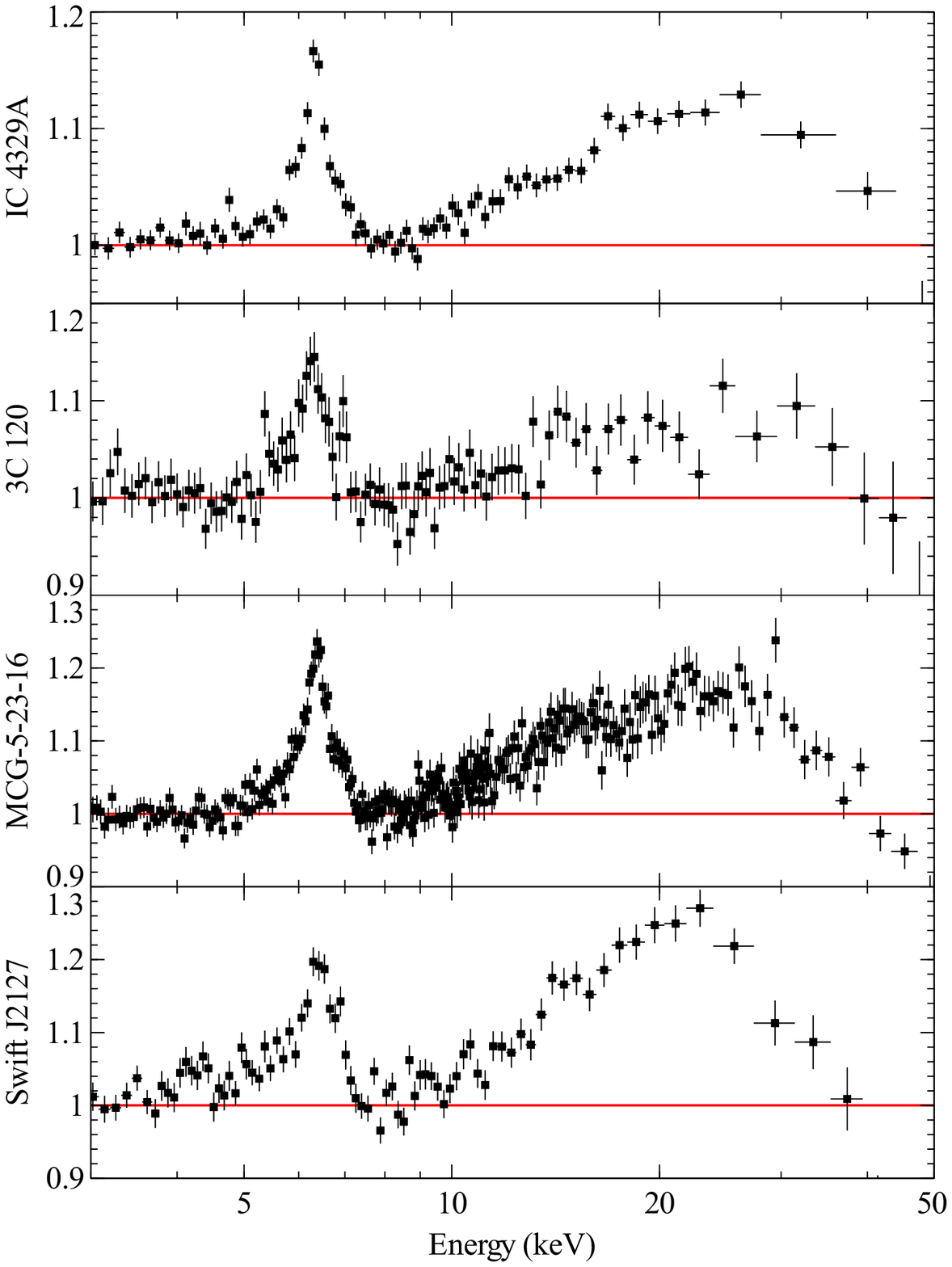}
\caption{NuSTAR spectra of a selection of AGN, shown as a ratio to a
  simple powerlaw fitted at 3-4 and 8-9 keV. Clockwise from top left
  are Mkn335, IC4329A, 3C120, MCG-5-23-16, SWIFTJ2127+5454,
  MCG--6-30-15, NGC4051 and NGC1365.  }
\end{figure*}

\section{NuSTAR Observations of Relativistic Lines}

Iron is the most abundant element with a high fluorescent yield, so
the iron-K emission line is a prominent feature of a typical reflection  
spectrum. Relativistic iron lines are common in AGN and in hard and
intermediate state BHB and have been observed with many different satellites and
detectors. Many observations show broad iron lines (see e.g. Miller
2007 and Fabian 2013 for reviews) and under the assumption that the
inner radius inferred from the width of the line corresponds to the ISCO,
black hole spin has been measured (See Brenneman in these Proceedings
and Reynolds 2013). 

Recently, NuSTAR has begun delivering high quality spectra from
3--78\,keV covering the iron line and Compton hump.  The spectrum of
NGC1365, for example, is well fitted by a relativistic reflection
model (Risaliti et al 2013).  Many other AGN now also show the
characteristic skewed iron line and Compton hump of reflection
(Fig.~4). Many of these objects are consistent with rapidly
spinning black holes, which leads to maximum blurrring. One AGN which
has a milder spin is SWIFTJ2127 shown in Fig.~3 from
Marinucci et al (2014).  With the first imaging hard X-ray telescope
for cosmic X-ray astronomy, NuSTAR gives excellent spectra of the
Compton hump to several tens keV which enables detailed modelling to
be carried out and the photon index of the primary continuum to be
measured with accuracy.

Broad iron lines have been confirmed with NuSTAR in several BHB
(GRO\,1915, Miller et al 2013; Cyg X-1, Tomsick et al 2013) and
even neutron stars (Ser X-1 Miller et al 2013). No distortion due to
pile up is present in NuSTAR spectra even in bright sources like these.
     
\begin{figure*}
\centering{
\psbox[xsize=7cm]
{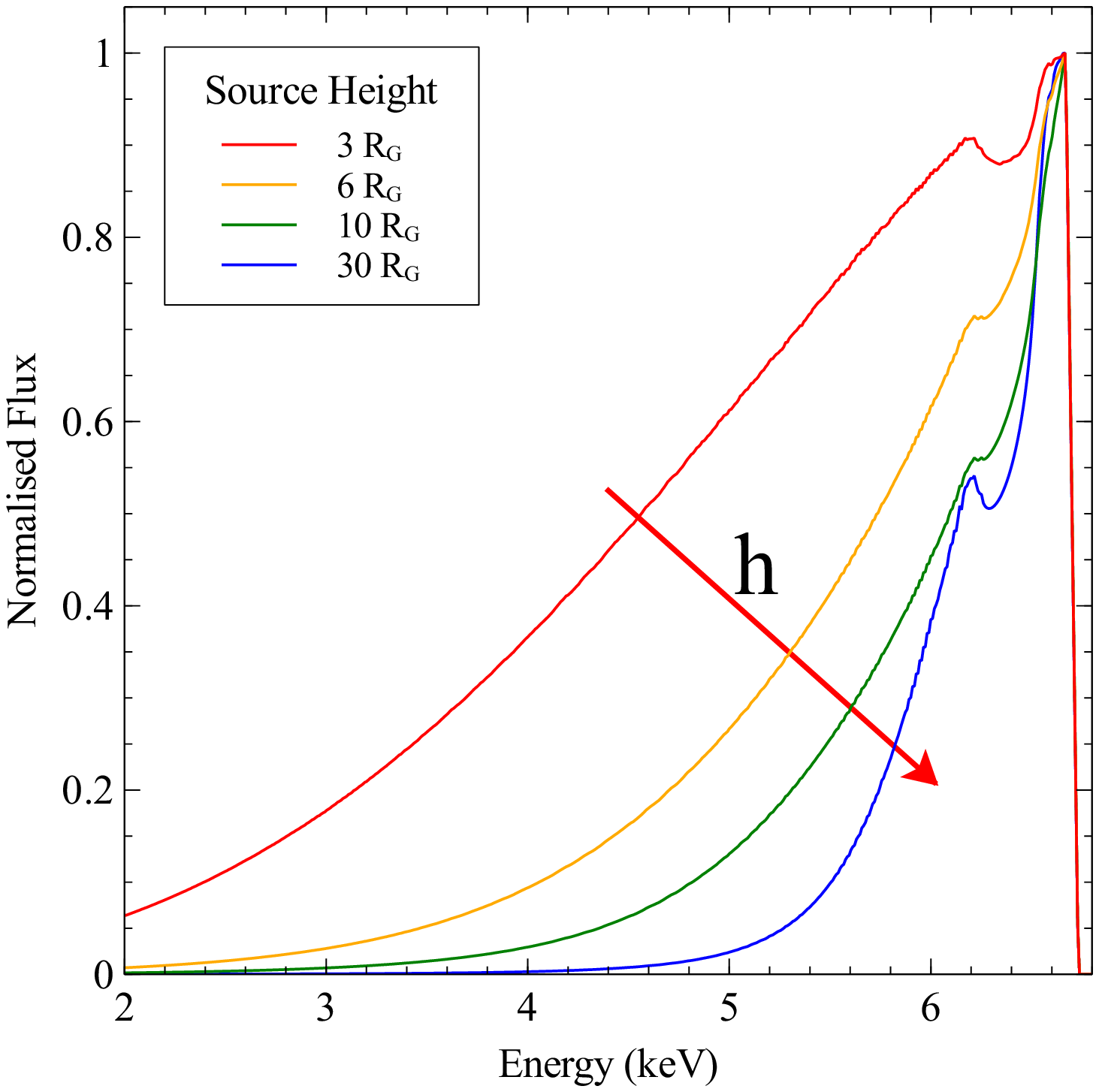}
\psbox[xsize=7cm]
{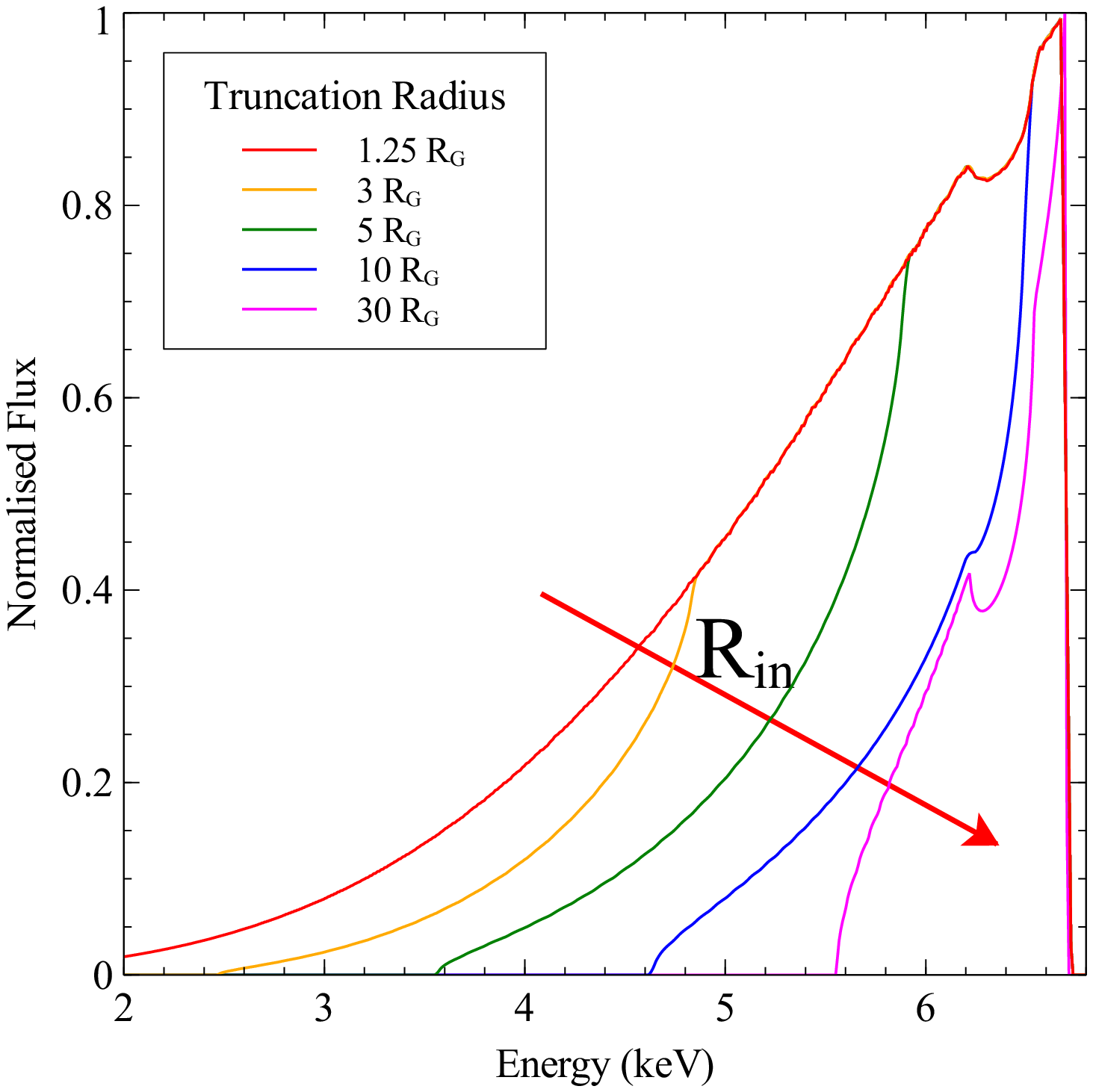}}
\caption{Expected broad iron-K line profiles for a lampost corona. 
On the  left the height of the corona is varied and on the right it is  
the inner  edge of the disc. The black hole is rapidly spinning in 
all cases (Fabian et al 2014).}
\end{figure*}

One point to note here is that a very broad line
implies not only that the black hole is spinning rapidly but also that
at least part of 
the corona is close to the black hole at a height less than 
$\sim10GM/c^2$ (Fabian et al 2014). If the corona is much further away
then it does not illuminate the innermost regions sufficiently for the
broad red wing to be clearly observed (see Fig.~5). Even if it is
closer, then if the matter in the corona is flowing outward at a mild
relativistic velocity, say at the base of a jet, then beaming can
reduce the illumination of the disc and thus the broad red wing.  

The corona does appear to be close to the black hole in many objects,
which then leads to the further relativistic effect of strong light
bending, Strong here implies a bending angle of up to a radian or
more. Light bending then has the effect of making an intrinsically
isotropic corona appear anisotropic to the outside
observer. Primary power-law radiation from the corona is bent toward the
disc (Fig.~1), enhancing the reflection spectrum and diminishing the observed
primary continuum. Some of the observed variability can then be due to
changes in source height (Miniutti \& Fabian 2004) rather than the
intrinsic luminosity. This can explain some of the puzzling
variability  seen in the AGN MCG--6-30-15 (Fabian et al 2003) and
may have been observed in XTEJ1650-500 (Rossi et al 2005; Reis et al 2013). 

Detailed modelling of the line shape in the best observed objects can
reveal the extent of the corona (Wilkins \& Fabian 2012) and its
changes during luminosity variations (Wilkins et al (2014). 

\begin{figure*}
\centering
\psbox[xsize=12cm]
{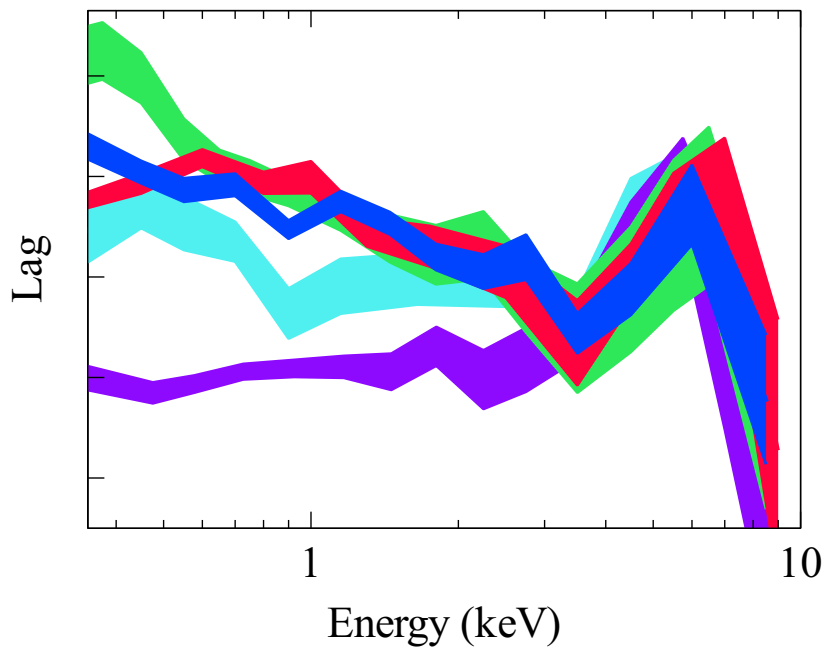}
\caption{The lag-energy spectra overplotted for five of the published
  sources with Fe K lags.  The amplitude of the lag has been scaled so
  that the lag between 3–4 keV and 6–7 keV match for all sources. The
  sources shown are: 1H0707-495 (blue), IRAS 13224-3809 (red), Ark 564
  (green), Mrk 335 (cyan) and PG 1244+026 (purple) (Kara et al
  2014). While the shape of the Fe K lags are similar in all these
  sources, the lags associated with the soft excess vary.}
\end{figure*}

\begin{figure*}
\centering
\psbox[xsize=10cm]
{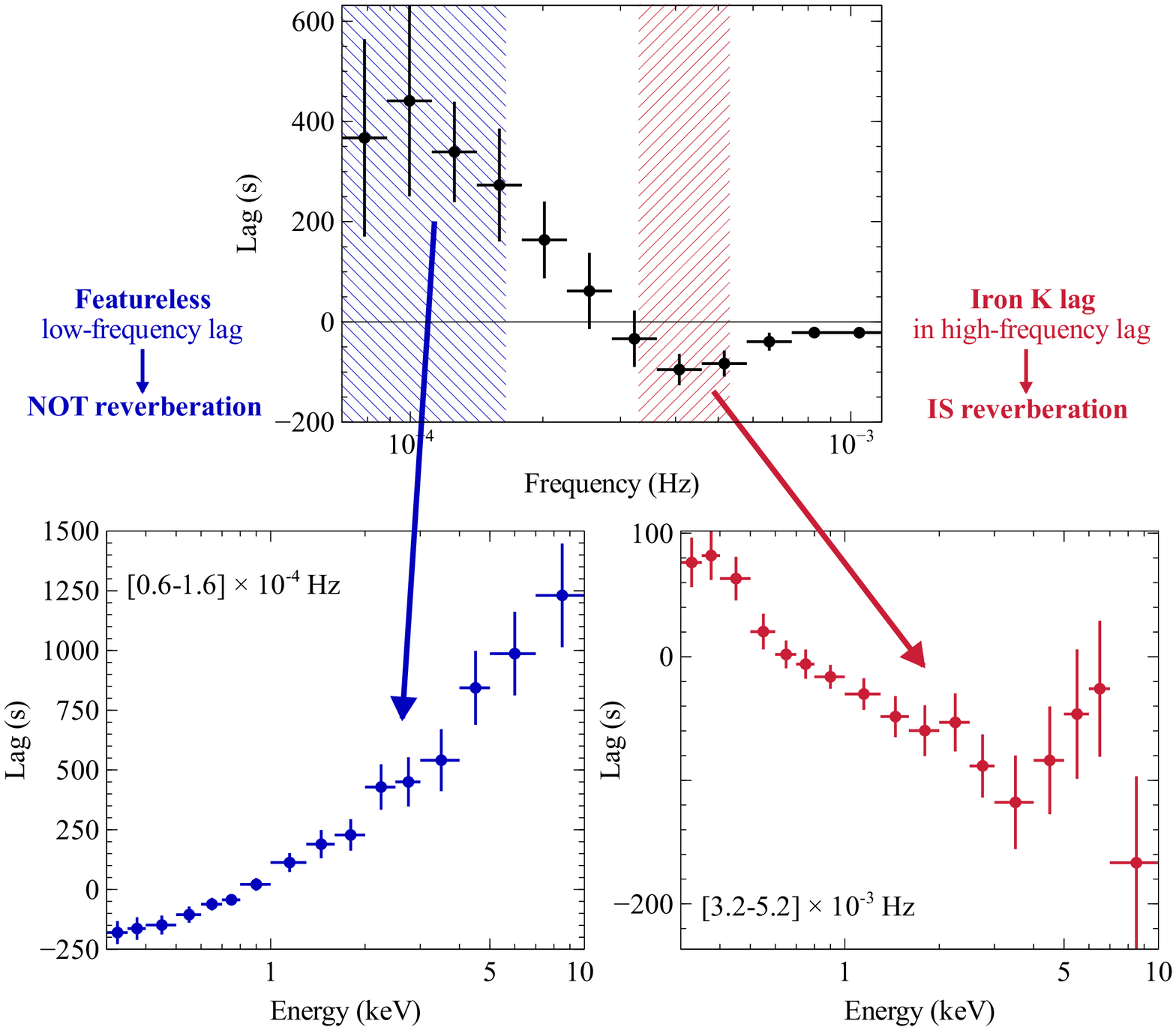}
\caption{Lag-frequency spectrum of Akn 564 is shown at the top, with
  the energy spectrum of the lags corresponding to the hatched2
  frequency ranges shown nelow. To the left is the featureless low
  frequency lag spectrum. The high frequency lag spectrum shows a
  clear iron line confirming  that it is due to reflection (Kara et al
  2013b). }
\end{figure*}

\section{Reverberation}

Soft X-ray reverberation has now been seen in about two dozen AGN (e.g.
De Marco et al 2013; Emmanoulopoulos et al 2010; de Marco et al 2011;
Kara et al 2013a; Alston et al 2013; Zoghbi \& Fabian 2011).  The 
reverberation timescales are of the
order of the light cross time of a few gravitational radii
(i.e.$GM/c^2$) implying that the corona is close to the black hole and
the disc extends close in, so the black hole must spin
rapidly. Iron-K reverberation is also now seen in about 9 AGN
(Fig.~6. Kara et al 2014; Cackett et al 2013; Uttley et al 2014).
Most of these results have been produced from the long continuous XMM light
curves that are possible due to its 48~hr orbit. The $\sim90$ min low
Earth orbits of
Suzaku and NuSTAR produce chopped light curves which makes timing analyses
challenging. Zoghbi et al (2013; 2014) have now overcome this problem with a
maximum likelihood method.  

\begin{figure*}
\centering
\psbox[xsize=12cm]
{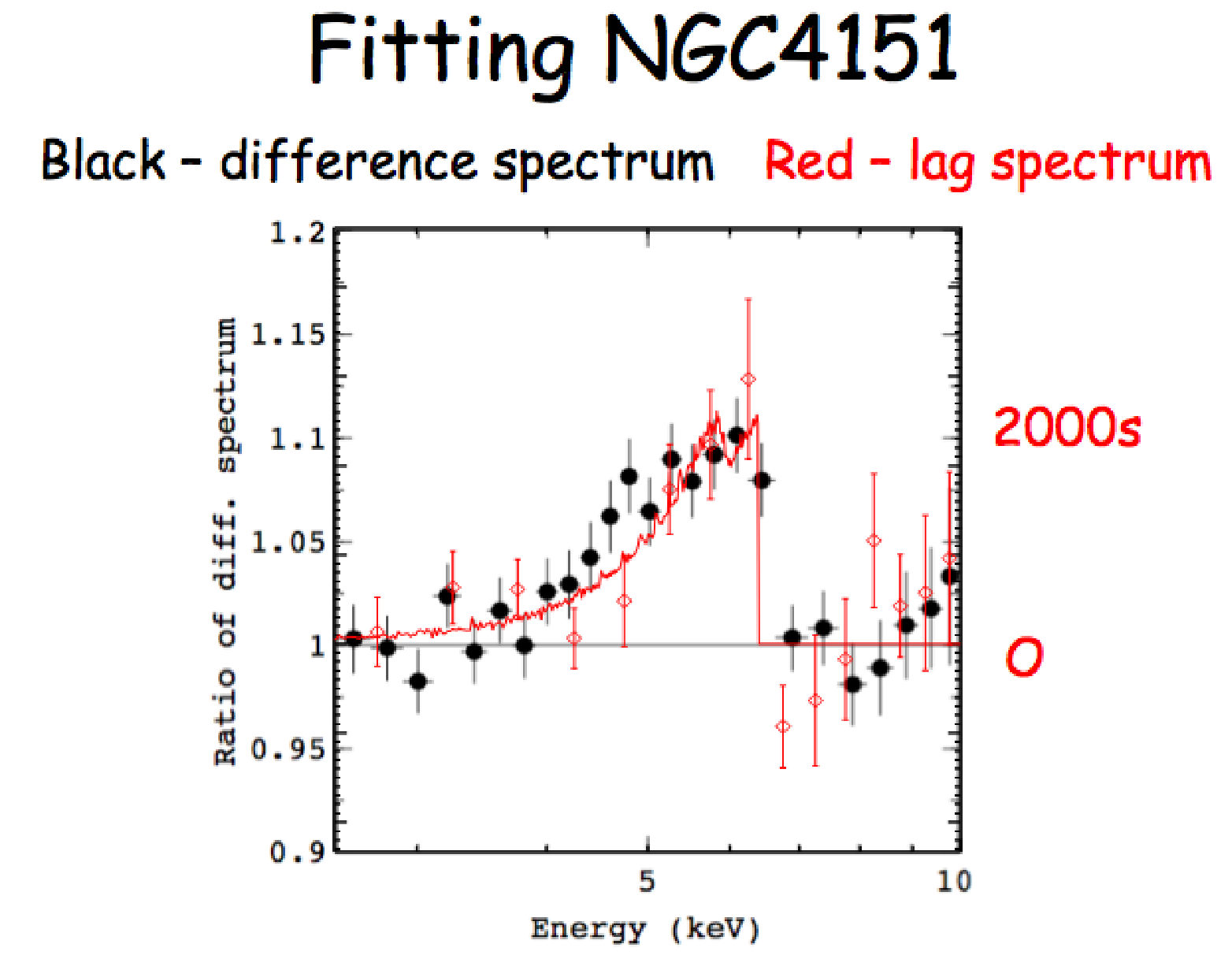}
\caption{The black points show the relativistic disc line spectrum in
  NGC4151 (left-hand axis, from a difference spectrum), whereas the
  red points show the lag-energy spectrum (right-hand axis). The
  agreement verifies that the black hole is rapidly spinning and has a
  mass close to 50 million \Msun (Cackett et al 2014).  }
\end{figure*}

Low frequency time lags were first seen in BHB such as Cyg X-1 by
Miyamato \& Kitamoto (1989) using GINGA data. They were observed at
frequencies $f<<10^{-2}t_{\rm cross}^{-1}$, where $t_{cross}=GM/c^3$
the light crossing time of the black hole. At these frequencies
propagation effects within the corona and changes in the corona, such
as variations in spectral index, dominate. Low frequency lags were also
seen in AGN (Vaughan et al. 2003; McHardy et al. 2004; Ar\'evalo et
al. 2006). The energy spectrum of
the lags is a relatively featureless power-law or steady increase with
energy (see also Walton et al 2013).

Reverberation dominates at {\em high} frequencies $f>10^{-2}t_{\rm
  cross}^{-1}$. Observation of a broad iron line in the high frequency
lags is a clear demonstration that they are due to reflection (Fig,~7;
Kara et al 2013). This is a very important confirmation of the
reflection scenario. Any alternative interpretation of the reflection
features (e.g. complex distant absorption, Turner \& Miller 2009; Miyakawa
et al 2012) now requires an explanation for the high frequency, iron-K
lags.

Modelling of the iron-K reverberation in NGC4151 (Fig.~8, Cackett et
al 2014) yields good agreement with its mass of $5\times 10^7 \Msun$,
measured by optical reverberation techniques.

\section{Discussion}
Relativistic disc lines are now seen in many luminous AGN and
BHB. Absence of broad lines in unobscured AGN can be due to the corona
being distant from the disc or outflowing at mildly relativistic
velocities. Generally, however, the height of the corona is
less than a few times the radius of the ISCO which means that the
effects of strong gravity due to GR are important. These include
gravitational redshifts, strong light bending and Shapiro time
delays (implicit in the reverberation lags). X-ray spectral-timimg analyses 
of several of the most rapidly
spinning black hole sources are giving us a window into the region
within one gravitational radius of the event horizon, the very
heart of the most luminous persistent sources of radiation in the Universe.

\section{Acknowledgements}

We thank the many colleagues who have contributed to the work reviewed
here, including Phil Uttley, Jon Miller, Ed Cackett, Dan Wilkins,
Chris Reynolds, Abdu Zoghbi, Dom Walton, Giorgio Matt, Fiona Harrison and the NuSTAR
team.

\section{References}
Alston, W. N., Vaughan, S.,  Uttley, P. 2013, MNRAS, 435, 1511\\
Cackett E et al 2013 ApJ, 720, L46 \\
Cackett E. et al 2014 MNRAS 438 2980\\
De Marco B et al 2011 MNRAS, 417, L98 \\
De Marco, B  et al. 2013, MNRAS, 431, 2441 \\
Emmanoulopoulos D. et al 2010, MNRAS, 404, 931\\
Fabian A.C. et al 1989 MNRAS, 238, 729 \\
Fabian AC Vaughan S., 2003, MNRAS, 340, L282\\
Fabian AC Ross RR 2010 SpaceSciRev 157, 167 \\
Fabian AC et al 2009 Nature, 459, 540 \\
Fabian AC 2013 IAUS, 290, 3\\
Fabian AC et al 2014 MNRAS, 439, 2307\\
Garcia J et al 2014 ApJ, 782, 76\\
Kara, E., et al 2013a, MNRAS, 428, 2795\\
Kara, E., et al 2013b, MNRAS, 434, 1129\\
Kara E., et al 2014, MNRAS, 439, L26\\
Marinucci A., et al 2014, 440, 2347\\ 
Miller JM, 2007 ARAA 45 441\\
Miller J.M et al 2013 ApJ, 799, L2\\
Miller J.M et al 2013 ApJ, 775, L45 \\
Miyakawa T., Ebisawa K., Inoue H., 2012, PASJ, 64, 140\\ 
Miyamoto S. Kitamoto S. 1989, Nature 342 773\\
Miniutti G., Fabian AC 2004 MNRAS, 349, 1435 \\
Reis R.C., et al 2013, ApJ, 763, 48\\
Reynolds CR 2013 CQGRa, 30, 4004\\
Rossi S., et al 2005, MNRAS, 360, 763\\
Risaliti G., et al 2013, Nature, 494, 449\\
Ross R.R., Fabian A.C., 1993, MNRAS, 264, 839\\
Ross R.R., Fabian A.C., 2005, MNRAS, 358, 211\\
Tanaka Y et al 1995 Nature 375 659 \\
Turner, T. J., Miller, L. 2009, A\&ARev, 17, 47\\
TomsickJ., et al 2014, ApJ, 780, 78\\
Uttley P. et al 2014, A\&ARev, in press\\
Walton DJ et al. 2013, ApJ, 777, L23\\
Wilkins D.R., Fabian A.C., 2012, MNRAS, 424, 1284\\
Wilkins D.R., Fabian A.C., 2013, MNRAS, 430, 247\\
Wilkins D.R., Fabian A.C., 2014, MNRAS, in press\\
Zoghbi A Fabian AC 2011 MNRAS, 418, 2642 \\
Zoghbi, A., et al 2012, MNRAS, 422, 1\\
Zoghbi A., et al 2013, ApJ, 777, 25 \\
Zoghbi A., et al 2014, ApJ, submitted\\ 

\label{last}

\end{document}